\documentstyle[twoside,fleqn,espcrc2]{article}
\def\beq{\begin{equation}}
\def\eeq{\end{equation}}
\def\bea{\begin{array}}
\def\eea{\end{array}}
\def\beqa{\begin{eqnarray}}
\def\eeqa{\end{eqnarray}}

\def\u1{{U(1)}}
\def\su2{{SU(2)}}
\newcommand{\re}{\relax{\rm I\kern-.18em R}}

\newcommand{\AmS}{{\protect\the\textfont2
  A\kern-.1667emt\lower.5ex\hbox{M}\kern-.125emS}}

\title{Lattice gauge theories and the Heisenberg antiferromagnetic chain
\thanks{Presented by 
G. Grignani at 
Lattice '99, Pisa, Italy.}
\vskip-3cm\hfill\small DFUPG-66-99\vskip2.6cm}
\author{F. Berruto, E. Coletti, G. Grignani and P. Sodano
\vspace{6pt}\\ {Dipartimento di Fisica and Sezione I.N.F.N.,
\vskip0.1cm Universit\'a di Perugia, Via Pascoli I-06123 Perugia, Italy}}
\begin{document}
\begin{abstract}
We study the strongly coupled 2-flavor lattice Schwinger model and the 
$SU(2)$-color $QCD_2$. The strong coupling limit, even with its
inherent nonuniversality, makes accurate predictions of the spectrum 
of the continuum models and provides an 
intuitive picture of the gauge theory vacuum. The massive excitations of the gauge model are 
computable in terms of spin-spin correlators of the quantum Heisenberg antiferromagnetic 
spin-1/2 chain.
\end{abstract}
\maketitle
\section{Introduction}

It is by now well known that, for lattice gauge theories, 
features of the strong coupling limit have analogs in quantum spin
systems.  In many cases, the problem of finding the vacuum 
of the strongly coupled lattice gauge theory is equivalent to
finding the ground state of a generalized quantum
antiferromagnet~\cite{smit}.  

Strong coupling limits of
gauge theories are highly non-universal: there are many choices of strong 
coupling theory which produce
identical continuum physics.  In spite of this difficulty, there exist
strong coupling computations which claim some degree of
success~\cite{noi}. In this contribution we review the results of the ground state and of the spectrum of the strongly coupled 2-flavor lattice Schwinger model and 
of $SU(2)$-color $QCD_2$.

We find that in order to determine the strong coupling vacuum, 
one has to solve
for the ground state of the Heisenberg antiferromagnetic spin chain, and 
that relevant quantities in the strong coupling expansion can be expressed 
in terms of spin-spin correlators of the quantum antiferromagnet.

\section{The 2-flavor Schwinger model}

The Hamiltonian and gauge constraint of the continuum 2-flavor Schwinger 
model are
\begin{equation}
H=\int dx[\frac{e^2}{2}E^2(x)+\psi^{\dagger}_a (x)\alpha\left(i\partial_x +eA(x)\right)\psi_a
(x)]\quad \label{h1}
\end{equation}
\begin{equation}
\partial_x E(x)+\psi^{\dagger}_a
(x)\psi_a (x)\sim 0\label{g1}\quad 
\end{equation}
with the flavor index $a$ taking the values $1,2$; there is summation 
over repeated indices. 
A lattice Hamiltonian reducing to (\ref{h1}) in the naive 
continuum limit is $H=H_0+\epsilon H_{h}$ with 
\begin{eqnarray}
H_0&=&\sum_{x=1}^{N}E_x^2\\
H_h&=&-i(R-L)\label{hop}
\end{eqnarray}
and $\epsilon=t/e^{2}a^{2}$.
In Eq.(\ref{hop}) the right $R$ and left $L$ hopping operators
($L=R^{\dagger}$) are 
$$
R=\sum_{x=1}^{N}R_{x}=\sum_{x=1}^{N}\sum_{a=1}^{2}R_{x}^{(a)}= 
\sum_{x=1}^{N} \sum_{a=1}^2 \psi_{a,x+1}^{\dag}e^{iA}\psi_{a,x}\quad .
$$
The lattice Gauss law constraint is given by
\begin{equation}
E_{x}-E_{x-1}+\psi_{1,x}^{\dag}\psi_{1,x}+\psi_{2,x}^{\dag}\psi_{2,x}-1\sim
0\ ,
\label{gauss}
\end{equation}
where the properly defined charge density reads
$
\rho(x)=\psi_{1,x}^{\dag}\psi_{1,x}+\psi_{2,x}^{\dag}\psi_{2,x}-1
$ and vanishes on every site occupied by only one
particle.

In a strong coupling perturbative expansion 
$H_0$ is the unperturbed Hamiltonian and $H_h$ the
perturbation; the ground state of $H_0$  is highly degenerate since
 each state with one particle per site has zero energy \cite{noi}.
 There are $2^N$ states of this type. First order perturbations to the vacuum energy vanish.
The leading term in the vacuum energy is  
of order $\epsilon^2$ 
\begin{equation}
E^{(2)}_{0}=<H_{h}^{\dagger}\frac{\Pi}{E_{0}^{(0)}-H_{0}}H_{h}>\quad ,
\label{secorder}
\end{equation}
where the expectation values are defined on the degenerate subspace
and $\Pi$ is a projection operator orthogonal to the set
of states with one particle per site. Since the charge density on the ground 
states of $H_{0}$ vanishes, one has that 
$
[H_0, H_h]=H_h
$
holds on any linear combination of the degenerate ground states. 
Consequently, from Eq.(\ref{secorder}), one finds
\begin{equation}
E^{(2)}_{0}=-2<RL>\quad .
\label{secorder2}
\end{equation}
Introducing the Schwinger spin operators
$
\vec{S}_{x}=\psi_{a,x}^{\dag}\frac{\vec{\sigma}_{ab}}{2}\psi_{b,x}
$
and taking into account that the Heisenberg Hamiltonian is 
$H_{J}=\sum_{x=1}^{N}(\vec{S}_x
\cdot \vec{S}_{x+1}-\frac{1}{4})$, on the degenerate subspace one has 
\begin{equation}
<H_{J}>=<\sum_{x=1}^{N}(-\frac{1}{2}L_{x}R_{x})>\quad ,
\label{mainequation}
\end{equation}
so that
\begin{equation}
E^{(2)}_{0}=4<H_{J}>\quad .
\label{secorder3}
\end{equation}
Finding the correct ground state amounts to the diagonalization 
of the Heisenberg spin 
Hamiltonian which is exactly diagonalizable in one dimension
\cite{bethe}. On a given site, the presence of a flavor $1$ particle 
can be represented, in the spin model, by the presence of a spin up, 
a flavor $2$ particle by the presence of a spin down. 
The number of the $H_J$ eigenstates is $2^N$.
Among these, the spin singlet with lowest energy is the non degenerate
ground state $|g.s.>$. Translational invariance of $|g.s.>$ 
amounts to the invariance of the gauge theory under the discrete chiral symmetry. The eigenvalue of the Heisenberg
Hamiltonian on this state in the thermodynamic limit is given by
$
H_{J}|g.s.>=(-N\  \ln\ 2)|g.s.>
$: due to Eq.(\ref{secorder3}) the second order correction to the vacuum energy is  $E_{g.s.}^{(2)}=-4 N\ln 2$.

There are two different types of excitations
which can be created from $|g.s.>$. Those that involve only spin
flipping and those that involve fermion transport besides spin
flipping.  The excitations of the first type have lower energy since
no electric flux is created, those of the second type have a higher
energy. The lowest energy ones occur when the fermion is
transported a minimal distance, since the energy is proportional to
the coupling times the length of the electric flux that is created.
Only the first type of excitations can be described in terms of the
Heisenberg model excited states.

In Ref\cite{noi} we showed that it is possible to identify the low lying 
excitations
of the Schwinger model with those of the Heisenberg model and that the
mass gaps of any other excitation can be expressed as functions of
v.e.v.'s of powers of $H_{J}$ and spin-spin correlation functions. We 
refer to \cite{noi} for a detailed analysis of the spectrum and of the 
chiral symmetry breaking pattern.

\section{The 2-color $QCD_2$}

The Hamiltonian and gauge constraint of the continuum $SU(2)$ color 
$QCD_2$ are 
\begin{eqnarray}
H&=&\int dx[\frac{e^2}{2}E^{a}(x)^2-i
\psi^{\dagger}_j (x)\alpha\partial_x \psi_j
(x)\nonumber\\
&-&\frac{g}{2}A^{a}(x)\psi^{\dagger}_{j}
(x)\alpha\sigma_{jk}^{a}\psi_{k}(x)]\nonumber\\ 
\partial_x E^a(x)&+&\epsilon^{abc}E^{b}(x)A^{c}(x)+\frac{1}{2}
\psi_{j}(x)^{\dagger}\sigma_{jk}^{a}\psi_{k}(x) \sim 0\nonumber
\end{eqnarray}
with $a,b,c=1,2,3$, $j,k=1,2$; there is summation over repeated indices. 
The lattice Hamiltonian is $H=H_0+\epsilon H_h$ with 
\begin{equation}
H_{0}=\sum_{x=1}^N\sum_{a=1}^{3} (E_{x}^{a})^2\label{eax}\ ,\quad
H_h=-i(R-L)\label{hop2}
\end{equation}
and $\epsilon=t/g^2a^2$. In Eq.(\ref{hop2}) the right $R$ and left $L$ 
hopping operators ($L=R^{\dagger}$) are
\begin{equation}
R=\sum_{x=1}^{N}R_x=\sum_{x=1}^N\sum_{j,k=1}^{2}
\psi_{j,x+1}
^{\dag}(e^{i\frac{{\vec \sigma}\cdot {\vec A}_{x}}{2}})_{jk}\psi_{k,x}
\end{equation}
The lattice Gauss law constraint reads
$$
E^a_{x}-e^{-i\frac{{\vec \sigma}\cdot {\vec A}_{x-1}}{2}}E^a_{x-1}
e^{i\frac{{\vec \sigma}\cdot {\vec A}_{x-1}}{2}}\nonumber\\
+\psi_{j,x}^{\dag}\frac{\sigma_{jk}^{a}}{2}\psi_{k,x}\sim
0\ ,
$$
and the charge density is given by $\rho_x=\psi_{1,x}^{\dagger}\psi_{1,x}+
\psi_{2,x}^{\dagger}\psi_{2,x}-1$. 

The ground state of $H_{0}$ is highly degenerate and the degeneracy 
equals $2^N$ since each site is either empty or occupied by two particles in a color singlet. First order perturbations to the vacuum energy vanish and the leading term of the vacuum energy is of order $\epsilon^2$ and it is given by  Eq.(\ref{secorder}) with $H_0$ and $H_h$ obtained from Eqs.(\ref{eax},\ref{hop2}) and $\Pi$ is the projection operator orthogonal to the degeneracy subspace. 

If one denotes by $C_2$ the quadratic Casimir in the fundamental representation of $SU(2)$, the equation 
$\left[H_0,H_h\right]=C_{2}H_h$ holds on any 
linear combination of the degenerate ground states. Following the same steps 
of section 2 and taking into account 
that the Schwinger spin operator is now given by 
$\vec{S}_x=\Psi^{\dagger}_x\frac{\vec{\sigma}}{2}\Psi_x$ where 
$\Psi=\left( \begin{array}{c}\psi_1\\ \psi_2^{\dagger}\end{array}\right)$,
 one gets 
\begin{equation}
E_{0}^{(2)}=\frac{4}{C_{2}}\langle H_{J}\rangle=\frac{16}{3}\langle H_{J}\rangle \quad   
\end{equation}
since $C_2=3/4$.

There are again two types of excitations created from the ground state: 
those involving only spin flipping (baryons) and those 
involving fermion transport besides spin flipping (mesons). 
Of course, the ones which do not involve charge transport have lower energy than the others. 
As we shall see shortly, the meson masses are computable in terms of 
spin-spin correlators of the Heisenberg Hamiltonian while baryons are 
the massless spinons of the spin-$1/2$ Heisenberg Hamiltonian.    
  
The lowest lying massive excitations are a pseudoscalar and a scalar meson 
created by the Fourier transform of the conserved gauge invariant currents 
at zero momentum $\sum_x j_1(x)=R+L$ and $\sum_x j_5(x)=R-L$, 
respectively; namely, 
\begin{equation}
|P\rangle=(R+L)|g.s.\rangle\label{pe}\ ,\ \ 
|S\rangle=(R-L)|g.s.\rangle\quad .\label{se}
\end{equation}
At the zero-th order they are degenerate, but the degeneracy is removed 
at the second order in the strong coupling expansion.
The mass of these excitations is obtained by computing their
energies  and by subtracting the ground state energy. 
Up to the second order in $\epsilon$, the mass of the state $|P\rangle$ 
is given by~\cite{erasmo}
\begin{eqnarray}
&m_P=\frac{g^2a}{2}(\frac{3}{4}-\epsilon^2\frac{20\langle 
\sum_{x=1}^N{\vec S_x}\cdot {\vec S}_{x+2}-\frac{1}{4}\rangle+16\langle H_{J}\rangle}
{3N+3\langle H_{J}\rangle})\nonumber \\
&=\frac{g^2a}{2}(\frac{3}{4}+\epsilon^2 54.10)\quad ,
\end{eqnarray}
whereas the one of the scalar meson $|S\rangle$ is
\begin{eqnarray}
&m_S=\frac{g^2a}{2}(\frac{3}{4}-\epsilon^2\frac{32\langle 
\sum_{x=1}^N{\vec S_x}\cdot {\vec S}_{x+2}-\frac{1}{4}\rangle-
128\langle H_{J}\rangle}
{2\langle H_{J}\rangle})\nonumber \\
&=\frac{g^2a}{2}(\frac{3}{4}+\epsilon^2 62.43)\quad .
\end{eqnarray}

\section{Concluding remarks}

We have shown how, in the strong coupling limit, all the massive low lying 
excitations of some gauge models can be computed in terms of spin-spin 
correlators of the Heisenberg model and that the massless 
excitations correspond to the 
spinons of the quantum antiferromagnet. As evidenced in 
\cite{korepin}, the explicit 
evaluation of the pertinent spin-spin correlators is far from 
being trivial and thus evaluating the spectrum in higher orders of 
strong coupling perturbative expansion may be quite involved. It is 
comforting to observe that, already at the second order in the strong coupling 
expansion, one gets results in satisfactory agreement with the continuum theory 
\cite{noi}.
For the $SU(2)$-color $QCD_2$, setting $t=1$, one gets $m_P=1.093$, which 
is $37\%$ higher than the result obtained in~\cite{bat} in the continuum for 
${\cal N}_c=2$ and is consistent with the lattice numerical calculations of ref.\cite{q8}.
For $m_S$ one gets $m_S=1.133$, this mass has not 
been computed in the continuum.

\end{document}